\documentclass[preprint,preprintnumbers]{revtex4}

\usepackage{graphicx}

\begin{document}

\title{Optical conductivity for  the surface of a Topological Insulator} 


\author{D. Schmeltzer$^{1}$}
\affiliation{$^{1}$Physics Department, City College of the City University of New York \\
New York, New York 10031, USA}

\author{K. Ziegler$^{2}$}
\affiliation{$^{2}$Institut f\"ur Physik, Universitat Augsburg  D-86135 Augsburg, Germany}

\begin{abstract}

The optical conductivity for the surface excitations for  a Topological Insulator as a function of the chemical potential and  disorder is  considered. Due to the time reversal symmetry   the chiral metallic  surface states are  protected against disorder. This allow to use the averaged  single particle Green's function  to compute  the optical conductivity.
We  compute the conductivity in the limit of a   finite disorder. We find that  the conductivity  as a function of the chemical potential $\mu $ and frequency $\Omega$ is given by the universal value  $\sigma(\Omega>2\mu)= \frac{e^2 \pi}{8h}$. For frequencies   $\Omega< \mu$   and elastic mean free path   $l_{el}=v\tau$  which obey   $k_{F}l>1$  we obtain the  conductivity   is given by   $\sigma(\Omega<2|\mu|)=\frac{e^2}{2h}\frac{k_{F}l_{el}}{(\Omega \tau)^2+1}$. In the limit of zero disorder  we find  $\sigma(\mu\neq0,\Omega ,k_{F}l\rightarrow \infty)=\frac{e^2 \pi}{2h}|\mu|\delta(\Omega)$.

\end{abstract}

\maketitle
\textbf{I. Introduction}

For  time reversal invariant systems one finds that for Kramer's states the time reversal operator $\vartheta$ obey $\vartheta^2=-1$ and thus the second Chern number is given by $(-1)^{\nu}=-1$ where $\nu$  is an odd number.
$Topological $ $Insulators$ $(TI)$ \cite{Volkov,Kane,Zhang,David}  are characterized by chiralic gapless electronic spectrum. For the  $3D$  $TI$ material  $Bi_{2}Se_{3}$ the surface state consist of a  $Weyl$  equation with single Dirac Cone  which is below the chemical potential $\mu$ and the  bulk gap $0.3eV$.  Due to the topology the backscattering is suppressed \cite{Balatzky} and therefore localization might  be prohibited.  Tunneling  scanning has  confirmed the presence of  backscattering and therefore the presence of a topological metallic state.  The conductivity results   are less conclusive, due the presence of the  $3D$  bulk gap  \cite{Culcer} or  the insulating  gap   for  thin layers $TI$.
The charge  current operator for  Weyl  equation  is identified with the spin on the surface of the $TI$. Therefore  for clean systems  conservation of momentum will cause    the low frequency conductivity to vanish, only the optical conductivity for frequency $\Omega>2\mu$  ($\mu$ is the chemical  potential) is finite.
In the presence of impurities  elastic scattering conserve energy but not the  quasi momentum,  giving rise to a finite Drude  conductivity.  Due to the spin orbit interaction anti-localization effects dominate the physics giving rise to a metallic conductivity \cite{Hikami,Ando,Stern,Shen,Hankiewicz}.
For the Weyl equation the velocity  operator is given   by:  $\frac{dr^{i}}{dt}=v_{F}\epsilon_{i,j}\sigma^{j}$and  ,$i,j=1,2$ and the charge current is given by the spin operator.  
We formulate the optical conductivity  for a finite chemical potential. We define   Green's  functions for  particles and anti-particles (holes).
Our results are as as following: For a  finite chemical potential   we find $\sigma(\Omega>2\mu)= \frac{e^2 \pi}{8h}[\mu+\frac{1}{4}f_{F.D.}(-\Omega+\mu)]$ which at low temperature  takes the universal value  $\sigma(\omega> \mu)=\frac{e^2 \pi}{8h}$.
The results obtained here in the absence of disorder are similar to the one obtained in  graphene \cite{Ziegler}.
For weak disorder   and chemical potential $\mu>\Omega$ we find the that the conductivity is given by the Drude  conductivity,  $\sigma(\Omega<2|\mu|)=\frac{e^2}{2h}\frac{k_{F}l_{el}}{(\Omega \tau)^2+1}$. ($l_{el}$ is the elastic mean free path).
The plan of this paper is as followings:
In section $II$ we introduce the model. In section $III$  we construct the Green's function for a model with a finite chemical potential. In chapter $IV$ we consider the effect of the white noise scattering potential on the single particles Green's functions.
In chapter $V$ we compute the conductivity in the single particle  approximation. Chapter $VI$ is devoted to conclusions.

\vspace{0.2 in}

\textbf{II- The Weyl model }

\vspace{0.2 in}

The surface state Hamiltonian  of a Topological Insulator ($TI$)of the  $Bi_{2}Se_{3}$ family materials  is given by a the Weyl model \cite{Raghou}.
The presence a  random potential and an  electromagnetic field  modify the Weyl model  in the following way:
\begin{eqnarray}
&&S=\int\,dt[\int\,d^2 r\Psi^{\dagger}(\vec{r},t)(i\partial_{t}-eA_{0}(\vec{r},t)\Psi(\vec{r},t)-H]; \nonumber\\&&H=\hbar v \int\,d^2 r \Psi^{\dagger}(\vec{r}) [\sigma^{1}(-i\partial_{2}-\frac{e}{\hbar}A_{2}(\vec{r},t))-\sigma^{2}(-i\partial_{1}-\frac{e}{\hbar}A_{1}(\vec{r},t))+ \frac{U_{sc}(\vec{r})}{{\hbar v}}]\Psi(\vec{r})\nonumber\\&&
\end{eqnarray}
$v$ is the Fermi velocity ,$\vec{A}(t)=\frac{i}{\Omega} \vec{E}(\Omega)e^{-i\Omega t}$
is the external  vector potential, $A_{0}(\vec{r},t)$ is the external scalar potential  and  $U_{sc}(\vec{r})$ is the random potential controlled by the   white noise correlation function  $<<U_{sc}(\vec{k}) U_{sc}(\vec{k}')>>=D_{sc}\delta^{2}[\vec{k}+\vec{k}']$. In the absence of the random potential and external fields  the  Hamiltonian (in the first quantization) in the momentum space  is  given by $h[\vec{k}]=-\sigma^{2}k_{1}+\sigma^{1}k_{2}$. The Hamiltonian $h[\vec{k}]$ is time reversal invariant  $\vartheta^{-1}h[\vec{k}]\vartheta h[-\vec{k}]$  ( $\vartheta=i\sigma^{2}K$ is the time reversal operator and $K$ is the conjugation operator which obey $\vartheta^2=-1$ )
This Hamiltonian has two eigenvectors : $|u^{(+)}(\vec{k})>$ with the eigenvalue $E_{p}=E_{+}=\hbar v |\vec{k}|$  for particles  and    $|u^{(-)}(\vec{k})>$ for  anti-particle $E_{a}=E_{-}=-\hbar v |\vec{k}|$.
\begin{eqnarray}
&&|u^{(+)}(\vec{k})>=\frac{1}{\sqrt{2}}|\vec{k}>\otimes[1,i e^{i\chi(k_{x},k_{y})}]^{T} \nonumber\\&&
|u^{(-)}(\vec{k})>=\frac{1}{\sqrt{2}}|\vec{k}>\otimes[-1,ie^{i\chi(k_{x},k_{y})}]^{T};
\chi(k_{x},k_{y})=tan^{-1}\left(\frac{k_{y}}{k_{x}}\right)
\end{eqnarray} 
The spinor operator $\Psi(\vec{r})$  is decompose into the eigen modes of the unperturbed  Weyl    Hamiltonian. 
\begin{equation}
\Psi(\vec{r})=\sum_{\vec{k}}e^{i\vec{k}\cdot \vec{r}}[C(\vec{k})u^{(+)}(\vec{k})+ B^{\dagger}(-\vec{k})u^{(-)}(-\vec{k})]\equiv\sum_{\vec{k}}e^{i\vec{k}\cdot \vec{r}}\Psi(\vec{k})
\label{gr}
\end{equation}
$\mu=\hbar v k_{F}$ stands for the chemical potential ,$|0>\equiv|\mu;p>\otimes|\mu;a>$. $|\mu;p>$ stands for particles ground state  and obeys   $C(\vec{k})|\mu;p>= 0$  for $ E(\vec{k})>\mu\equiv \hbar v k_{F} >0$ ($k_{F}$  is the Fermi momentum. For $\mu<0$  we have in the ground state antiparticle for energies $E(\vec{k})< |\mu|$). $|\mu;a>$ represent the antiparticles ground state and obeys  $B(\vec{k})|\mu;a>=0$.
Using the eigen spinors  the Hamiltonian is equivalent to two coupled bands.
\begin{eqnarray}
&&H_{0}=\sum_{\vec{k}}[\hbar v|\vec{k}|C^{\dagger}(\vec{k})C(\vec{k})+ \hbar v|\vec{k}|B^{\dagger}(-\vec{k})B(-\vec{k})]\nonumber\\&&
H_{D}=\sum_{\vec{k}}\sum_{\vec{p}} U_{sc}(\vec{k}-\vec{p})\Big[ V^{p,p}(\vec{k}, \vec{p})C^{\dagger}(\vec{k})C(\vec{p})+V^{a,a}(-\vec{k}, -\vec{p})B(-\vec{k})B^{\dagger}(-\vec{p})\nonumber\\&& +
V^{p,a}(\vec{k}, -\vec{p}) C^{\dagger}(\vec{k})B^{\dagger}(-\vec{p})+V^{a,p}(-\vec{k}, \vec{p})B(-\vec{k})C(\vec{p})\Big]\nonumber\\&&
\end{eqnarray} 
$H_{0}$ is the unperturbed Hamiltonian and $H_{D}$ is the effect of the random potential on the two bands.
The spinors structure gives rise to  vertex functions for the coupling to the random potential in momentum space The vertex functions are given in terms of the particles and anti-particles matrix elements where  $p$ stands for particles and  $a$ stands for antiparticles:  are given by:
\begin{eqnarray}
&& V^{p,p}(\vec{k}, \vec{p})=<u^{(+)}(\vec{k})|u^{(+)}(\vec{p})>;  V^{a,a}(-\vec{k}, -\vec{p})=<u^{(-)}(-\vec{k}))|u^{(-)}(-\vec{p})>;\nonumber\\&&
 V^{p,a}(\vec{k}, -\vec{p})=<u^{(+)}(\vec{k})|u^{(-)}(-\vec{p})> ; V^{a,p}(-\vec{k}, \vec{p})=<u^{(-)}(-\vec{k})|u^{(+)}(\vec{p})> \nonumber\\&&
\end{eqnarray}
  
In agreement with the time reversal invariance $\vartheta^2=-1$  we find  that the backscattering is prohibited:
$V^{p,p}(\vec{k},\vec{p}=-\vec{k})= e^{\frac{i}{2}(\chi(\vec{k})-\chi(-\vec{k}))}\cos[\frac{1}{2}(\chi(\vec{k})-\chi(-\vec{k}))]$. From the representation  $\chi(k_{1},k_{2})=\chi (\vec{k})=tan^{-1}(\frac{k_{2}}{k_{1}})$  we obtain the relation $ \chi(-\vec{k})=\pi+  \chi(\vec{k})$ and find that $V^{p,p}(\vec{k},\vec{p}=-\vec{k})=0$.

\vspace{0.2 in}

\textbf{III-The single particle Green's function  for a finite chemical potential $\mu$ in the  absence of disorder }

\vspace{0.2 in}

In  the absence of disorder we  can use  the spinor operator   $\Psi_{\sigma}(\vec{k})= [ \Psi_{\sigma=\uparrow}(\vec{k}), \Psi_{\sigma=\downarrow}(\vec{k})^{T}]$ with the components :
\begin{equation}
\Psi_{\sigma}(\vec{k})=C(\vec{k})u_{\sigma}^{(+)}(\vec{k})+ B^{\dagger}(-\vec{k})u_{\sigma}^{(-)}(-\vec{k})
\label{eq}
\end{equation}
Using the spin half spinor $|\sigma=\pm\frac{1}{2}\rangle$  we obtain the the projected  spinors   $u_{\sigma}^{(+)}(\vec{k})$  and $u_{\sigma}^{(-)}(\vec{k}) $ 
\begin{equation}
u_{\sigma}^{(+)}(\vec{k})=\langle\sigma|u^{(+)}(\vec{k})\rangle,\hspace{0.1 in} u_{\sigma}^{(-)}(\vec{k})=\langle\sigma|u^{(+)}(\vec{k})\rangle
\label{spi}
\end{equation}
We define   the  single particle Green's  matrix  $ ig_{\sigma,\sigma'}(\vec{k},\vec{k}';t,t')= i\delta_{\vec{k},\vec{k}'}G_{\sigma,\sigma'}(\vec{k};t-t'=\tau)$ \cite{Abrikosov}.
\begin{eqnarray}
&&iG_{\sigma,\sigma'}(\vec{k},\tau)=<0|T(\Psi_{\sigma}(\vec{k},\tau)\Psi_{\sigma'}^{\dagger}(\vec{k},0))|0>=  (u_{\sigma}^{(+)}(\vec{k}))^*\cdot u_{\sigma'}^{(+)}(\vec{k})<\mu;p|T(C(\vec{k},\tau)C^{\dagger}(\vec{k},0))|\mu;p>+\nonumber\\&&(u_{\sigma}^{(-)}(-\vec{k}))^*\cdot u_{\sigma'}^{(-)}(-\vec{k})<\mu;a|T(B^{\dagger}(-\vec{k},0)B(-\vec{k},\tau))|\mu;a>\equiv\nonumber\\&&  iG^{p}(\vec{k},\tau)(u_{\sigma}^{(+)}(\vec{k}))^*\cdot u_{\sigma'}^{(+)}(\vec{k})+iG^{a}(-\vec{k},\tau) (u_{\sigma}^{(-)}(-\vec{k}))^*\cdot u_{\sigma'}^{(-)}(-\vec{k})\nonumber\\&&
\end{eqnarray}
The Fourier  transform in the frequency domain allows to write: $G^{p}(\vec{k},\omega= \hat{\omega}+\mu)$ for particles and $G^{a}(-\vec{k},\omega= \hat{\omega}+\mu)$ for antiparticles ( $p$  stand for particles and  $a$ for anti-particles Green's function).
 The Green's functions for a positive  chemical potential   $\mu>0$ are given by in terms of the Fermi Dirac function   for temperatures $T\rightarrow 0$, $ f_{F.D.}(\hat{\omega})=\frac{1}{e^{\beta(\hat{\omega})}+1};\beta\rightarrow \infty$
\begin{eqnarray}
&&G^{p}(\vec{k},\omega=\hat{\omega}+\mu;\mu>0)=\frac{1-f_{F.D.}(\hat{\omega})}{\hat{\omega}- (E(\vec{k})-\mu))+i\epsilon}+ \frac{f_{F.D.}(\hat{\omega})}{ (E(\vec{k})-\mu)-i\epsilon}, \epsilon\rightarrow 0 \nonumber\\&&
G^{a}(-\vec{k},\omega;\mu>0)=\frac{1-f_{F.D.}(\hat{\omega}+\mu)}{\hat{\omega}+ E(\vec{k})+\mu-i\epsilon}, \epsilon\rightarrow 0\nonumber\\&&
\end{eqnarray}

\vspace{0.2 in}

\textbf{IV-The single particle Green's function in the presence  of the white noise potential $U_{sc}(\vec{r})$ }

\vspace{0.2 in}

To second order in the scattering potential $U_{sc}(\vec{r})$ the self energy for the particles is given by  $\Sigma(\vec{k},\omega;p)$ for the anti-particles the self energy $\Sigma(\vec{k},\omega;a)$ is given by a similar result:
\begin{eqnarray}
&&\Sigma(\vec{k},\omega;p)=\int\,\frac{d^2 k' }{(2\pi)^2}<<U_{sc}(\vec{k}-\vec{k}')U_{sc}(\vec{k}'-\vec{k})>> V^{p,p}(\vec{k},\vec{k}') V^{p,p}(\vec{k}',\vec{k})G^{(p)}(\vec{k}',\omega)\nonumber\\&&
+ \int\,\frac{d^2 k' }{(2\pi)^2}<<U_{sc}(\vec{k}-\vec{k}')U_{sc}(\vec{k}'-\vec{k})>> V^{p,a}(\vec{k},-\vec{k}') V^{a,p}(-\vec{k}',\vec{k})G^{(a)}(-\vec{k}',\omega)\nonumber\\&&
\Sigma(\vec{k},\omega;p)\equiv \Sigma_{R}(\vec{k},\omega;p)+i \Sigma_{I}(\vec{k},\omega;p) \nonumber\\&&
\Sigma_{R}(\vec{k},\omega;p=\hat{\omega}+\mu,\Lambda;p)=\Gamma^{2}_{\|}\int_{0}^{ v \Lambda}\,\frac{d^{2}E}{\hat{\omega}-(E-\mu)}+ \Gamma^{2}_{\perp}\int_{0}^{ v \Lambda}\,\frac{d^{2}E}{\hat{\omega}+(E+\mu)}\nonumber\\&&
 \Sigma_{I}(\vec{k},\omega;p)=  -\pi\Gamma^{2}_{\|}(\hat{\omega}+\mu)  [f_{F.D.}(-\hat{\omega})- f_{F.D.}(\hat{\omega})] -\pi\Gamma^{2}_{\perp}(\hat{\omega}+\mu)f_{F.D.}(-\hat{\omega}-\mu) \nonumber\\&&
\Gamma^{2}_{\|}\approx \Gamma^{2}_{\perp} 
\approx \frac{ D_{sc}}{4\pi  v^2} ;
\end{eqnarray}
This allows to define the life time $\tau$, $ \frac{1}{2\tau}=\frac{D_{sc}\mu}{4v^2}$.
The notation $<<...>>$ stands for the white noise average and $\Lambda$ is the momentum cut-off.
The real part of the self energy diverges in the limit $\omega=0$.
 The derivative of the self energy introduces the  \textbf{wave function renormalization $Z^{-1}$   (or the quasi particle weight)}:
$Z^{-1}(\hat{\omega},\lambda;\Lambda) =[1-\partial_{\hat{\omega}}\Sigma_{R}(\hat{\omega},\mu,\lambda,\Lambda;p)|_{\hat{\omega}=0}]$
As a result we replace the chemical potential $\mu$ by  the renormalized  chemical potential:$\mu\rightarrow \mu=Z[\mu- \Sigma_{R}(0,\mu,\lambda,\Lambda;p))]$
Similarly the  velocity $v$ is replaced by : $v\rightarrow v Z(\hat{\omega},\lambda;\Lambda)$
The averaged Green's functions   for particles  $\overline{G^{p}}(\vec{k};\omega)$ and anti-particles   $\overline{D^{a}}(\vec{k};\omega)$ is given by :
$\overline{G^{p}}(\vec{k};\omega)=\int_{0}^{\infty}\,ds$ $\overline{G^{p}}(\vec{k};s=t-t')e^{i\omega s}+ \int_{-\infty}^{0}\,ds \overline{G^{p}}(\vec{k};s=t-t')e^{i\omega s}\equiv \overline{G^{p}}(\vec{k};\omega,+)+\overline{G^{p}}(\vec{k};\omega,-)$  and  
$\overline{D^{a}}(\vec{k};\omega)=\int_{0}^{\infty}\,ds \overline{G^{a}}(\vec{k};s=t-t')e^{i\omega s}+\int_{-\infty}^{0}\,ds \overline{G^{a}}(\vec{k};s=t-t')e^{i\omega s}\equiv \overline{G^{a}}(\vec{k};\omega,+)+\overline{G^{a}}(\vec{k};\omega,-)$.
 The averaged  Green's functions can be expressed  in terms of the spectral functions $A_{\pm}$ for particles and $B_{\pm}$  for antiparticles .
\begin{eqnarray}
&&\overline{G^{p}}(\vec{k};\omega,+)\equiv \int_{-\infty}^{\infty}\,dz\frac{A_{+}(\vec{k},z)}{\hat{\omega}-(E(\vec{k})-\mu)+i\epsilon}; G^{p}(\vec{k};\omega,-)\equiv \int_{-\infty}^{\infty}\,dz\frac{-A_{-}(\vec{k},z)}{\hat{\omega}-(E(\vec{k})-\mu)-i\epsilon}
\nonumber\\&&
A_{+}(\vec{k},\hat{\omega})= (\frac{1}{4\pi \tau})\frac{ (1-f_{F.D.}(\hat{\omega}))}{(\hat{\omega}-(E(\vec{k})-\mu))^2+(\frac{1}{2\tau})^2};
A_{-}(\vec{k},\hat{\omega})= (\frac{1}{4\pi \tau})\frac{  f_{F.D.}(\hat{\omega})}{(\hat{\omega}-(E(\vec{k})-\mu))^2+(\frac{1}{2\tau})^2};\nonumber\\&&
\overline{G^{a}}(\vec{k};\omega,+)\equiv \int_{-\infty}^{\infty}\,dz\frac{B_{+}(\vec{k},z)}{\hat{\omega}+(E(\vec{k})+\mu)+i\epsilon}; G^{a}(\vec{k};\omega,-)\equiv \int_{-\infty}^{\infty}\,dz\frac{-B_{-}(\vec{k},z)}{\hat{\omega}+(E(\vec{k})+\mu)-i\epsilon}
\nonumber\\&&
B_{+}(\vec{k},\hat{\omega})= (\frac{1}{4\pi \tau})\frac{ (1-f_{F.D.}(\hat{\omega}))}{(\hat{\omega}+(E(\vec{k})+\mu))^2+(\frac{1}{2\tau})^2};
B_{-}(\vec{k},\hat{\omega})=(\frac{1}{4\pi \tau})\frac{ f_{F.D.}(\hat{\omega})}{(\hat{\omega}+(E(\vec{k})+\mu))^2+(\frac{1}{2\tau})^2};\nonumber\\&&
\end{eqnarray}

\vspace{0.2 in}

\textbf{V-Computation of the current in the averaged single particle  approximation}

\vspace{0.2 in}

We replace the ground state $ |0>$ with the average effective  ground state $|\phi_{0}>$ characterized by the spectral functions $A_{\pm}$ and $B_{\pm}$.
We construct the evolution operator due to  the external potential $H^{ext}(t)$ which acts on the effective ground state  $|\phi_{0}>$. We use the interaction picture and compute 
the induced current  $\delta J_{1}(\vec{r},t)$ to linear order in the vector potential $\vec{A}(t)$  (see eq.$1$):
The  Hamiltonian which describe  the coupling of light to matter is given by :
\begin{equation}
H^{ext}(t)=\int\,d^{2}r [J_{1}(\vec{r},t)A_{1}(\vec{r},t)+J_{2}(\vec{r},t)A_{2}(\vec{r},t)]
\label{eq}
\end{equation}  
We use the interaction picture and compute 
the induced current  $\delta J_{1}(\vec{r},t)$ by  the vector potential $\vec{A}(t)$  (see eq.$1$):
\begin{equation}
\delta J_{1}(\vec{r},t)=\frac{-i}{\hbar}\int_{-\infty}^{t} dt'<\phi_{0}|[\hat{J}_{1}(\vec{r},t),H^{ext}(t')]|\phi_{0}>
\label{equation}
\end{equation}
The current operator is defined by the variation of the Hamiltonian with respect the vector potential:
\begin{eqnarray}
&&\hat{J}_{1}(\vec{r},t)\equiv\frac{\delta H^{ext}(t)}{\delta A_{1}(\vec{r},t)}=(-e)v\Psi^{\dagger}(\vec{r})\sigma^{2}\Psi(\vec{r})\nonumber\\&& 
\hat{J}_{2}(\vec{r},t) \equiv\frac{\delta H^{ext}(t)}{\delta A_{2}(\vec{r},t)}=(-e)v\Psi^{\dagger}(\vec{r})(-\sigma^{1})\Psi(\vec{r})\nonumber\\&&
\end{eqnarray} 
we obtain    the linear response for  the conductivity  $ \sigma_{1,1}(\vec{q},\Omega)$ \cite{Doniach}.
\begin{eqnarray}
&&\sigma_{1,1}(\vec{q},\Omega)=\frac{R_{1,1}(\vec{q},\Omega)}{i\Omega}\nonumber\\&&
R_{1,1}(\vec{q},\Omega)=\int_{-\infty}^{\infty}\,ds e^{is\Omega} R_{1,1}(\vec{q},s) \nonumber\\&&
R_{1,1}(\vec{q},s=t-t')=\frac{-i}{\hbar}\theta[t-t']<\phi_{0}|[\hat{J}_{1}(\vec{q},t),\hat{J}_{1}(-\vec{q},t')]||\phi_{0}> \nonumber\\&&
\end{eqnarray}
$\theta[t-t']$ is the step function which is one for $t\geq t'$.
The current operator  $\hat{J}_{1}(\vec{r})$ is build from the four components  $\hat{J}^{p,p}_{1}(\vec{q};t)$,$\hat{J}^{a,a}_{1}(\vec{q};t)$,$\hat{J}^{a,p}_{1}(\vec{q};t)$ $\hat{J}^{p,a}_{1}(\vec{q};t)$ and the spinor matrix elements :
$W_{1}^{p,p}( \vec{k}+\vec{q},\vec{k})=<u^{(+)}(\vec{k}+\vec{q})|\sigma^{2}|u^{(+)}(\vec{k})>$;$ W_{1}^{a,a}( -\vec{k}-\vec{q},-\vec{k})=<u^{(-)}(-\vec{k}-\vec{q})|\sigma^{2}|u^{(-)}(-\vec{k})>$ ;$ W_{1}^{a,p}( -\vec{k}-\vec{q},\vec{k})=<u^{(-)}(-\vec{k}-\vec{q})|\sigma^{2}|u^{(+)}(-\vec{k})>$; $ W_{1}^{p,a}( \vec{k}+\vec{q},-\vec{k})=<u^{(+)}(\vec{k}+\vec{q})|\sigma^{2}|u^{(-)}(-\vec{k})>$
\begin{eqnarray}
&& \hat{J}^{p,p}_{1}(\vec{q};t)=\int\,\frac{d^2k}{(2\pi)^2}C^{\dagger}(\vec{k}+\vec{q},t)
W_{1}^{p,p}( \vec{k}+\vec{q},\vec{k})C(\vec{k},t)\nonumber\\&&
\hat{J}^{a,a}_{1}(\vec{q};t)=\int\,\frac{d^2k}{(2\pi)^2}B(-\vec{k}-\vec{q},t)
W_{1}^{a,a}( -\vec{k},-\vec{q},-\vec{k})B^{\dagger}(-\vec{k},t)\nonumber\\&&
\hat{J}^{p,a}_{1}(\vec{q};t)=\int\,\frac{d^2k}{(2\pi)^2}C^{\dagger}(\vec{k}+\vec{q},t)
W_{1}^{p,a}( \vec{k}+\vec{q},-\vec{k})B^{\dagger}(-\vec{k},t)\nonumber\\&&
\hat{J}^{a,p}_{1}(\vec{q};t)=\int\,\frac{d^2k}{(2\pi)^2}B(-\vec{k}-\vec{q},t)
W_{1}^{a,p}( -\vec{k}-\vec{q},\vec{k})C(\vec{k},t)\nonumber\\&&
\end{eqnarray}
We can express the conductivity in terms of the four currents build from the particles $p$ and anti-particles $a$ :$\hat{J}^{p,p}_{1}(\vec{q};t)$ ,$\hat{J}^{p,a}_{1}(\vec{q};t)$ $\hat{J}^{p,a}_{1}(\vec{q};t)$, $\hat{J}^{a,a}_{1}(\vec{q};t)$. 
\begin{eqnarray}
&&\sigma_{1,1}(\vec{q},\Omega)=\frac{R_{1,1}(\vec{q},\Omega;p,p;p,p)+R_{1,1}(\vec{q},\Omega;p,a;a,p)+R_{1,1}(\vec{q},\Omega;a,p;p,a)+R_{1,1}(\vec{q},\Omega;a,a;a,a)}{i\Omega}\nonumber\\&&
\end{eqnarray}
Using the explicit form  of  the spectral functions $A_{\pm}$,$B_{\pm}$ given in equation $(10)$ we obtain the conductivity.

\textbf{a) The conductivity in the limit  of infinitesimal  disorder with a vanishing chemical potential} 
 
The only finite  contributions are  given by the combination $B_{-}A_{+}$ , the other two  spectral functions  $B_{+}$ and $A_{-}$ are zero.
\begin{eqnarray}
&&\sigma_{1,1}(\Omega;\mu\rightarrow 0)=\frac{e^2 \pi}{2h}\int_{-\infty}^{\infty}\,d\omega\int_{0}^{\Lambda}E\, dE[\frac{B_{-}(E,\omega)A_{+}(E,\omega+\Omega)-B_{-}(E,\omega+\Omega)A_{+}(E,\omega)]}{\Omega}]=\nonumber\\&&
\frac{e^2 \pi}{2h}\int_{-\infty}^{\infty}\,d\omega\int_{0}^{\Lambda}E\, dE[\frac{f_{F.D.}(\hat{\omega})(1-f_{F.D.}(\hat{\omega}+\Omega))}{\Omega}\delta(\hat{\omega}+E)\delta(\hat{\omega}+\Omega-E) \nonumber\\&&- \frac{f_{F.D.}(\hat{\omega}+\Omega)(1-f_{F.D.}(\hat{\omega}))}{\Omega} \delta(\hat{\omega} +\Omega+E)\delta(\hat{\omega}-E)]=\nonumber\\&&
\frac{e^2 \pi}{8h}[ f_{F.D.}(-\frac{\Omega}{2})(1-f_{F.D.}(\frac{\Omega}{2})) + f_{F.D.}(\frac{\Omega}{2})(1-f_{F.D.}(-\frac{\Omega}{2})]|_{T\rightarrow 0}\rightarrow \frac{e^2 \pi}{8h}\nonumber\\&&
\end{eqnarray}

\textbf{b)-The conductivity for a finite chemical potential  $\mu>0$ in the limit  of infinitesimal  disorder }

For this case $B_{+}=0$ and we have three nonzero spectral functions ,$A_{+}$,$A_{-}$and $
B_{-}$. 
\begin{eqnarray}
&&\sigma_{1,1}(\Omega;\mu>0)=\frac{e^2\pi}{2h}\int_{-\infty}^{\infty}\,d\omega\int_{0}^{\Lambda}E\, dE[ (\frac{-A_{-}(E,\omega)A_{+}(E,\omega-\Omega)+A_{+}(E,\omega)A_{-}(E,\omega-\Omega)}{\Omega}) \nonumber\\&&+(\frac{-B_{-}(E,\omega+\Omega)A_{+}(E,\omega)+B_{-}(E,\omega)A_{+}(E,\omega+\Omega)}{\Omega})]=
\nonumber\\&&
\frac{e^2\pi}{2h}[\mu_{R}\delta(\Omega)+\frac{1}{4}f_{f.D.}(\frac{\Omega}{2})f_{f.D.}(-\frac{\Omega}{2}-\mu)+\frac{1}{4}(1-f_{f.D.}(\frac{\Omega}{2}))(1-f_{f.D.}(\frac{\Omega}{2}-\mu_{R}))]|_{T\rightarrow 0}\nonumber\\&&\rightarrow \frac{e^2\pi}{2h}[\mu\delta(\Omega)+ \frac{1}{4}f_{F.D.}(-\frac{\Omega}{2}+\mu)]\nonumber\\&&
\end{eqnarray}
We find that the the conductance at  $T=0$ is given by  the universal value $\frac{e^2 \pi}{8h}$ for $\Omega>2\mu$.   For a finite chemical potential the metallic behavior  is given by  $\frac{e^2\pi}{2h}|\mu|\delta(\Omega)$. 
In figure $1$ we show the conductivity for the entire  range of frequencies  for the case that the elastic mean free path is large.

\textbf{c)- The conductivity in the limit  of \textbf{finite   disorder}   with the  finite life time case $l_{el.}=v \tau$ at a low frequency  $\Omega< 2\mu$.}

In the limit of in the limit $\Omega\rightarrow 0$ we can limit ourself only to the conductivity of the conduction band  which is given by given by $R_{1,1}(0,\Omega;p,p;p,p)$. We  find Drude like behavior given  by  figure $2$.
\begin{equation}
\sigma_{1,1}(\Omega<2|\mu|)=\frac{e^2}{2h}\frac{k_{F}l_{tr}}{(\Omega \tau)^2+1}
\label{cond}
\end{equation}
Where $k_{F}$ is given by $k_{F}=\frac{\mu}{\hbar v}$. We observe that the conductivity contains the factor half.  The origin of the factor half is due to the angle dependent vertex  $\cos^{2}[\chi(\vec{k})]$. In addition we remark that Ladder correction will replace the scattering time $\tau$ and therefore the elastic mean free path  with the transport time $\tau_{tr} > \tau $  and therefore  $l$ with $l_{tr}>l$.
In figure $2$ we have plotted the  conductivity for this case.

\textbf{Conclusion}

To conclude we have computed the optical conductivity for the Weyl Hamiltonian which describe the surface excitations of the $TI$ for the entire  range of frequencies. Due to the time reversal invariance $\vartheta^2=-1$  the backscattering is prohibited  and the model belongs to the sympectic ensemble justifying the use of the single particle  approximation.
The universal value of the conductivity is given by   $\sigma(\omega> \mu)=\frac{e^2 \pi}{8h}$. For  a finite chemical potential 
and finite disorder we find  that the conductivity is given by  $\sigma(\Omega<2|\mu|)=\frac{e^2}{2h}\frac{k_{F}l_{tr}}{(\Omega \tau)^2+1}$.
In the limit of  $\mu\approx 0$  self consistent calculations performed by one of us \cite{Ziegler2} show that the conductivity is given by $\sigma(\mu\approx 0,\Omega=0)\approx \frac{e^2}{\pi h}$ 
\begin{figure}
\begin{center}
\includegraphics[width=7.25 in ]{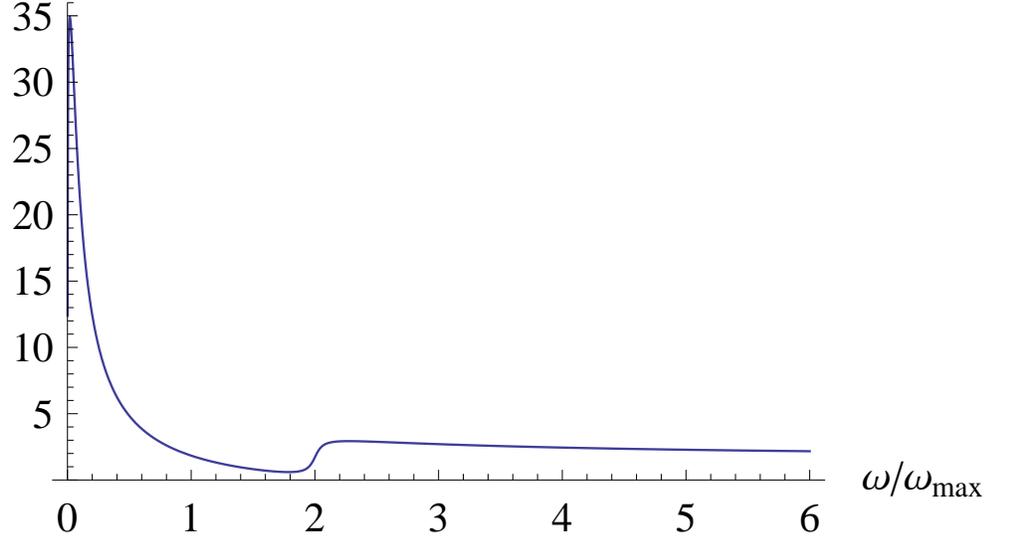}
\end{center}
\caption{The conductivity for the entire  frequency} 
\end{figure}

\begin{figure}
\begin{center}
\includegraphics[width=7.25 in ]{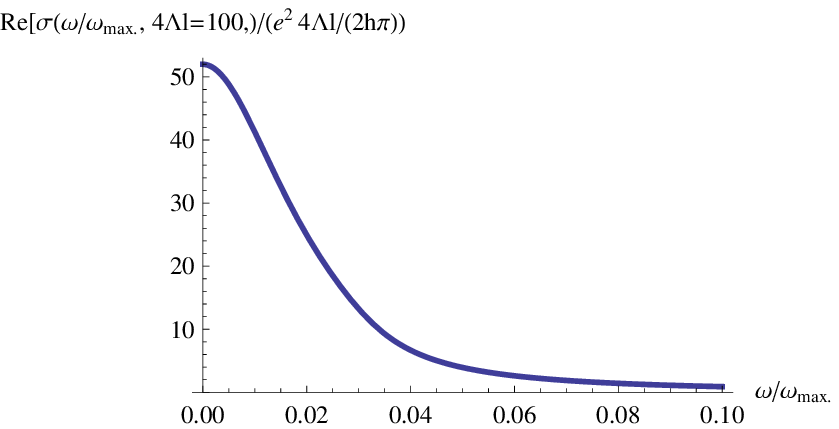}
\end{center}
\caption{The conductivity at low frequencies} 
\end{figure}




\end{document}